\documentclass[aps,prl,floatfix,twocolumn,preprintnumbers,nofootinbib,reprint,groupedaddress,amssymb,superscriptaddress]{revtex4-1}
\PassOptionsToPackage{dvipsnames}{xcolor}

\usepackage{amsmath}
\usepackage{amsfonts}
\usepackage{amssymb}
\usepackage{comment}
\usepackage{orcidlink}
\usepackage{mathtools}
\usepackage{braket}
\usepackage{graphicx}
\usepackage{rotating}
\usepackage{latexsym}
\usepackage{array}
\usepackage{xcolor}
\usepackage{multirow}
\usepackage{array,makecell}
\usepackage[utf8]{inputenc}
\usepackage{mathrsfs}
\usepackage{tabularx}
\usepackage{slashed}
\usepackage{hyperref}
\usepackage{booktabs}
 
\hypersetup{colorlinks, citecolor=bluscuro, linkcolor=bluscuro, urlcolor=bluscuro}
\definecolor{rossos}{cmyk}{0,1,1,0.55}
\definecolor{bluscuro}{rgb}{0.15, 0.2, .85}
\definecolor{bluchiaro}{cmyk}{1,.3,0.,0.1}
\definecolor{verdescuro}{rgb}{0.3,0.8,0.3}

\newcommand{\be}{\begin{equation}}
\newcommand{\ee}{\end{equation}}
\newcommand{\bea}{\begin{eqnarray}}
\newcommand{\eea}{\end{eqnarray}}
\newcommand{\bc}{\begin{center}}
\newcommand{\ec}{\end{center}}

\newcommand{\GeV}{\,\mathrm{GeV}}

\def\cale{{\mathcal{E}}}

\def\MG{\texttt{MadGraph}}
\def\PT{\textsc{Pythia} 8}

\begin{document}

\title{
Bump Hunting Inside Jets with Energy Correlators
}

\preprint{CERN-TH-2026-116}

\author{Lorenzo Ricci~\orcidlink{0000-0001-8704-3545}}
\affiliation{Maryland Center for Fundamental Physics, Department of Physics, University of Maryland, College Park, MD 20742, USA}

\author{Marc Riembau~\orcidlink{0000-0002-9842-2425}}
\affiliation{Theoretical Physics Department, CERN, 1211 Geneva 23, Switzerland}
\affiliation{Theoretical Particle Physics Laboratory, Institute of Physics, EPFL, Lausanne, Switzerland}

\author{Minho Son~\orcidlink{0000-0001-9913-7784}}
\affiliation{Department of Physics, Korea Advanced Institute of Science and Technology,
	291 Daehak-ro, Yuseong-gu, Daejeon 34141, Republic of Korea}

\begin{abstract}
\noindent
{

Energy correlators exhibit well-understood scaling behavior in the collinear 
limit, governed by perturbative QCD dynamics. We explore how this scaling 
regime is broken by new physics, converting precise energy correlator measurements into a broadband search for new physics. Under generic assumptions, unitarity and positivity are sufficient to classify and constrain the relevant signatures, which imprint an angular resonance on top of this smoothly scaling background. This converts the search into bump hunting within jets.  As a proof of principle, we derive projected LHC sensitivity for a light hadrophilic $Z'$, finding competitive 
constraints with existing searches.

}
\end{abstract}

\maketitle

\section{Introduction}

Jets are information-rich objects copiously produced at the LHC, whose internal 
structure encodes both perturbative and nonperturbative aspects of QCD dynamics. 
Their characterization is certainly one of the central themes of modern collider phenomenology.

A broad class of Beyond the Standard Model (BSM) scenarios predict jet-like 
signatures generated by dynamics other than QCD, and therefore with a 
dramatically different internal structure. Classifying these scenarios 
theoretically and identifying observables capable of discriminating them from 
ordinary QCD jets is an active area of 
research~\cite{Strassler:2006im,Kaplan:2008pt,Beauchesne:2018myj,Bernreuther:2019pfb,
Cheng:2021kjg,Knapen:2021eip,Born:2023vll,Cheng:2024hvq,Schwaller:2015gea,McCullough:2022hzr,
Batz:2023zef,Cohen:2023mya,Carrasco:2023loy,Albouy:2022cin}.
The theoretical and experimental toolkit for probing exotic jet structure has grown substantially, encompassing jet substructure variables, machine learning discriminants, and model-agnostic anomaly detection 
methods~\cite{Larkoski:2017jix,Kogler:2018hem,Nachman:2020ccu,Kogler:2021kkw,Karagiorgi:2022qnh,Belis:2023mqs}. Many of these approaches, 
however, are sensitive to accurate modeling of parton showers and hadronization, and typically lack sharp, first-principles and theoretically robust features to anchor the searches.

Energy Correlators (ECs) are a particularly clean class of jet substructure observables, originally proposed at LEP~\cite{Basham:1978zq}. 
Recently, they have regained substantial attention through new theoretical developments and are now actively explored in LHC studies. 
They measure the average energy deposited along specific directions on the detector surface as a 
function of the angles around the collision point. This framework has attracted 
considerable attention across a wide range of SM applications (see, 
e.g.,~\cite{Moult:2025nhu} for a recent overview). Most prominently, the 
collinear limit of two detector operators at small angular separations 
exhibits a characteristic power-law scaling~\cite{Komiske:2022enw}, which has 
recently enabled a precise CMS determination of the strong coupling 
constant~\cite{Chen:2023zlx,CMS:2024mlf}.

In this \textit{Letter} we propose ECs as a complementary approach to searching for new physics. Any new hard scale introduced by BSM dynamics explicitly breaks the scaling behavior described above. In particular, we show that a broad class of BSM scenarios, characterized by a boosted resonance $X$ of mass $m_X$ decaying hadronically, produces a sharp, resonant excess in the ECs at a fixed angular scale $z_\star \sim m_X^2/E_J^2$, where $E_J$ is the jet energy.

This converts the new physics search into bump hunting in angular space, with a theoretically predicted spectral shape encoding the spin and polarization of the resonance $X$. We further show that new physics ``shapes'' are bounded by unitarity and energy positivity once the resonance mass $m_X$ is fixed. This approach is largely model-independent and carries a distinct set of theoretical and experimental systematics compared to conventional jet substructure searches. We further explore this idea by studying sensitivity projections for a light hadrophilic $Z'$ at the LHC.

\section{Angular resonances inside jets}\label{Sec:GenBSM}

ECs provide a systematic probe of the angular distribution 
of energy flow inside jets, theoretically framed as expectation values of energy 
flow operators $\mathcal{E}_n$ \cite{Sveshnikov:1995vi,Hofman:2008ar,Belitsky:2013xxa,Kravchuk:2018htv}. The two-point energy correlator (EEC) $\langle\mathcal{E}_{n_1}\mathcal{E}_{n_2}\rangle$ 
measures the expected product of energy fluxes through two calorimeters at directions 
$\vec{n}_1$ and $\vec{n}_2$, and is typically reported as a function of their angular 
separation $z = \frac{1-\vec{n}_1\cdot \vec{n}_2}{2}$. In the collinear regime $z\ll 1$, 
it develops a scaling behavior $\langle\mathcal{E}_{n_1}\mathcal{E}_{n_2}\rangle \sim \frac{1}{z}\bigl(1 + \gamma\log z + \cdots\bigr)$
where $\gamma$ is an anomalous dimension calculable in perturbative QCD. In traditional 
QCD language this scaling regime arises from DGLAP evolution of collinear radiation; 
it has recently been reinterpreted in terms of the light-ray OPE of the energy 
detectors~\cite{Hofman:2008ar}. The scaling behavior receives power corrections~\cite{Korchemsky:1999kt,Chen:2024nyc,Lee:2024esz} 
suppressed by $z_\text{QCD}/z$, where $z_\text{QCD}\sim \Lambda_\text{QCD}/E_J$ 
and $E_J$ is the jet energy; only at sufficiently small angular scales $z\sim z_\text{QCD}$, 
confinement induces order-one corrections. Any additional hard scale explicitly breaks 
this scaling behavior, and it is precisely this handle that we exploit.

We consider BSM scenarios where a fraction of jets is initiated not by quarks 
or gluons, but by a boosted particle $X$ decaying hadronically. Rather than 
targeting a specific model, we can work within a space of models defined by two 
broad assumptions. First, $i)$ $X$ couples linearly to a QCD operator 
$\mathcal{O}$ ($\mathcal{L}\supset g_X X \mathcal{O}$), so that the energy 
correlator sourced by $X$ takes the form
\begin{equation}\label{eq:twopointcorr}
\langle\mathcal{E}_{n_1}\mathcal{E}_{n_2}\rangle = \frac{1}{\mathcal{N}} 
\int d^4x\, e^{iq\cdot x}\langle 0|\, \mathcal{O}^\dagger(x)\, 
\mathcal{E}_{n_1}\mathcal{E}_{n_2}\, \mathcal{O}(0)\,|0\rangle\,,
\end{equation}
where $q^\mu$ is the four-momentum of $X$ and $\mathcal{N}$ is the normalization by the total rate. Second, $ii)$ $X$ is sufficiently 
narrow, $\Gamma_X/m_X\ll 1$, so that it is dominantly produced on-shell, 
$q^2\simeq m_X^2$. The EC is then convoluted with the Breit-Wigner distribution,  
$\sim m\Gamma/((q^2-m_X^2)^2+m_X^2\Gamma_X^2)$, which in the narrow-width limit 
reduces to a delta function localizing $q^2$ to $m_X^2$. Under these two 
assumptions, once $\mathcal{O}$ is specified, the signal depends only on 
$m_X$ and $g_X$.

\medskip
\textit{\textbf{Unitarity-bounded classification.---}}
The quantum numbers of $\mathcal{O}$ determine the shape of the ECs. We 
classify operators by the number of elementary QCD fields they contain and 
by their spin. Bilinear operators admit a complete classification, which we 
carry out here; operators with three or more fields give rise to multi-prong 
topologies. We leave them to future work.

For a bilinear operator, the EEC of \eqref{eq:twopointcorr} in the rest frame, as a function of $z$ and for $z>0$,\footnote{There is of course another contribution at $z=0$, namely when the detectors are on top of each other. However, the phenomenologically interesting case is at finite separation.} is just a delta function in the back-to-back configuration, $\delta(1-z)$. Indeed, the decays are back-to-back and the energy flows through opposite directions. QCD corrections resolve the delta function and are discussed below. When the operator $\mathcal O$ has spin $J$ and $X$ is produced in a polarized state, the correlator acquires a nontrivial dependence on $\cos\theta$, the angle between the detector axis $\vec n_1-\vec n_2$ and the spin axis of $X$. The EEC at Born level can be written as a sum over even Legendre polynomials 
\begin{align}\label{Eq:PVD}
    \braket{\cale_{n_1}\cale_{n_2}} \,=\, \frac12 \delta(1-z) \left( 1+\sum_{j=2}^{2J} c^{(j)}P_j(\cos\theta) \right)\,,
\end{align}
where $J$ is the spin of $\mathcal O$, $P_j$ Legendre polynomials and the sum runs over even $j$. 
The shape is therefore fully characterized by the spinning coefficients $c^{(j)}$. As explained in \cite{Riembau:2025isw}, these encode both the helicity state of $X$ and the spin projection of the QCD partons on the detector axis $\vec n_1-\vec n_2$. Unitarity of the underlying theory together with energy positivity of the detector operators bound the allowed range of $c^{(j)}$, so that the space of admissible signals is compact, and in fact, convex.

 For a spin-1 source decaying to two QCD particles, a single coefficient $c^{(2)}$ characterizes any possible model, and it is constrained to $-1 \leq c^{(2)}\leq \frac{1}{2}$~\cite{Hofman:2008ar}. The allowed parameter space is convex, meaning that any hypothetical model consistent with the assumptions generates a signal which is a combination of the extreme cases, i.e. $\text{EEC}(z)=t\, \text{EEC}_{c^{(2)}=1/2} + (1-t)\text{EEC}_{c^{(2)}=-1}$ for $0\leq t \leq 1$. Examples of operators saturating both
 boundaries for a transversely polarized vector are shown in Table~\ref{tab:operators_spin}.
For a spin-2 source, the parameter space is two-dimensional, spanned by $c^{(2)}$ and $c^{(4)}$ and bounded by three inequalities $1+c^{(2)}+c^{(4)}\geq 0$, $1-\frac12 c^{(2)}-4c^{(4)}\geq 0$ and $1-c^{(2)}+6c^{(4)}\geq 0$ forming a bounded triangle. 
Representative examples saturating vertices of the triangle are shown in Table~\ref{tab:operators_spin}. Each example in Table~\ref{tab:operators_spin} can be realized by a particular model such as axion-like particles (ALPs), $Z^\prime$, KK-gravitons and so on. 

\begin{table}[]
        \renewcommand{\arraystretch}{1.50} 
        \addtolength{\tabcolsep}{11.00pt}
	\begin{tabular}{|c|c|l|}
  \hline
	  Spin & Operator  & Spinning coeff. \\ \hline \hline
	  \multirow{2}{*}{0}    & $q\bar{q}$ & $-$ \\
	 	    & $\text{tr}\left (G_{\mu\nu}G^{\mu\nu}\right )$ & $-$ \\ \hline
          \multirow{2}{*}{1}    & $\bar q \gamma^\mu q$ & $ c^{(2)}=1/2$ \\
	 	    & $\bar q \sigma^{\mu\nu}\partial_\mu  q$ &  $c^{(2)}= -1$ \\ \hline
          \multirow{3}{*}{2}    & $\bar q \gamma^{\mu}\partial^{\nu} q$ & $(c^{(2)},c^{(4)})=(-\frac57,-\frac{2}{7})$\\
	 	    & $\bar q \sigma^{\mu\rho}\partial_\rho\partial^\nu q$ &  $(c^{(2)},c^{(4)})=(-\frac{10}{7},\frac{3}{7})$ \\
		    & $\mathrm{tr}(G_{\mu\rho}G^{\rho}{}_{\nu})$ & $(c^{(2)},c^{(4)})=(\frac{10}{7},\frac{1}{14})$ \\ \hline         
	\end{tabular}
	\caption{Bilinear QCD singlet operators and their associated spinning coefficients.}
	\label{tab:operators_spin}
\end{table}

In the rest frame, integrating over $\cos\theta$ washes out the sensitivity to the spinning coefficients $c^{(j)}$ and therefore the spin information, so the EEC becomes universal across spins and polarizations. However, the same does not hold in the boosted frame, and the EEC functional form does depend on the spin of $\mathcal O$ and the polarization of $X$, as was shown in \cite{Ricci:2022htc} for the electroweak boson case.

The boost is characterized by an angular scale $z_\star\equiv m_X^2/E^2$, where $m_X$ is the particle mass and $E$ the total jet energy. For helicity eigenstates the boost axis coincides with the spin quantization axis of $X$, and each detector direction is characterized by a polar angle $\cos\theta_i$ with respect to it. In the back-to-back limit, we have $\cos\theta_1\simeq - \cos\theta_2$, and the geometry is described by a single angle $\theta_1$. Under the boost, $n_i^\mu = (1,\vec{n}_i)$ transforms to $n_i'^\mu$ from which the detector in the frame where $X$ is boosted is obtained by the rescaling $n_i'^\mu = \lambda_i(1,\vec{n}_i')$ with $\lambda_i = (1+ \sqrt{1-z_\star} \cos\theta_i)/\sqrt{z_\star}$~\cite{Belitsky:2013bja,Riembau:2025isw}. The polar angles $\theta_i'$ in the boosted frame are given by $\cos\theta_i' = \frac{\cos\theta_i + \sqrt{1-z_\star} }{1+ \cos\theta_i \sqrt{1-z_\star}}$. This is just the relativistic aberration formula with $\beta=\sqrt{1-z_\star}$.  
The angular separation between detectors in the boosted frame $z^\prime$ depends not only on the separation $z$ in the rest frame, but also on their individual locations with respect the spin axis, and it is given by $z = \lambda_1 \lambda_2 z'$. 
The back-to-back configuration of particles in the rest frame with the delta-function $\delta(1-z)$ localizes the detector orientation to $\cos^2\theta_1=\frac{1-z_\star/z^\prime}{1-z_\star}$, which has a solution for $z^\prime > z_\star$. Together with the phase space Jacobian which includes a factor $1/\sqrt{1-z_\star/z'}$, this leads to a sharp resonance peak at $z^\prime/z_\star\simeq 1$, with an edge singularity that is resolved by finite width and the QCD corrections. 

\begin{figure}
	\centering
	\includegraphics[width=.90\columnwidth]{"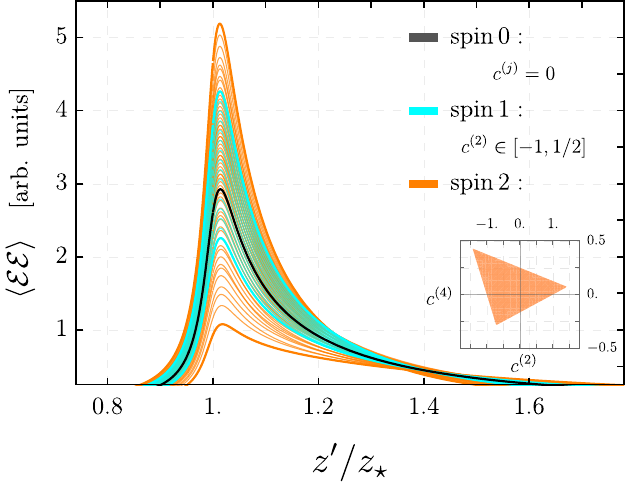"}
	\caption{Spectrum of two-point energy correlators allowed by unitarity for a spin 0, 1 and 2 resonance. The multiple resonances for spin 1 and 2 correspond to discretely selected spinning coefficients.}
	\label{fig:resonantshape}
\end{figure}

The spinning coefficients $c^{(j)}$ control the polar angle distribution of the decay products, and since those relate to $z^\prime$ through the aberration, the spinning information in the boosted frame is encoded in the functional form in terms of $z^\prime$.
While different spins and polarizations produce peaks at the same position $z_\star$, distinct spectral shapes of resonances are expected, as shown in Fig.~\ref{fig:resonantshape}. Remarkably, this makes the observable discriminating beyond a simple mass measurement.

The Born-level distributions for a spin 0 and a transversely polarized spin 1 and 2 as allowed by unitarity are shown in Fig.~\ref{fig:resonantshape}. In Fig.~\ref{fig:resonantshape}, we fix $z_\star=0.2$ and we assume a width similar to the SM $Z$-boson, $\Gamma_X/m_X\simeq \Gamma_Z/m_Z$. The relevant kinematic variable is $z^\prime/z_\star$, which indeed shows a resonant behavior at $z^\prime/z_\star\simeq 1$ regardless of the spin, but with spectral shapes that strongly depend on both the spin and the spinning coefficient $c^{(j)}$. In the spin-1 case, the extremal correlators with the highest and lowest peak correspond to $c^{(2)}=-1$ and $c^{(2)}=1/2$, respectively. In the spin-2 case, they correspond to the $(c^{(2)},c^{(4)})=(-10/7,3/7)$ and $(10/7,1/14)$ points, respectively. Any model within our two assumptions produces a correlator that is a convex combination of the boundary values of its spin, so Fig.~\ref{fig:resonantshape} maps out the space of allowed signals for spins up to 2. Given the difficulties of having higher spin particles in isolation (see, e.g., \cite{Bellazzini:2019bzh}), Fig.~\ref{fig:resonantshape} describes the entire space of consistent models. 

The peak at $z^\prime/z_\star\simeq 1$ is the most distinctive feature of this class of models and is robust against QCD corrections. QCD radiation primarily affects the spectrum at $z^\prime/z_\star\lesssim 1$, and resolving the Born-level delta function into a smooth distribution (see, e.g., \cite{Ricci:2022htc, Holguin:2026vld,Gao:2026xuq}), while leaving the localized threshold enhancement intact. For $z^\prime/z_\star\ll 1$, the correlator reduces to the standard QCD correlator on a hard parton. The search strategy therefore focuses on $z^\prime\simeq z_\star$, where the resonant localized excess sits above the smooth QCD background. Its position encodes $m_X$ and its functional form encodes the spin and polarization of $X$.

\section{Hunting for a hadrophilic $Z^\prime$}\label{Sec:Pheno}

As a first concrete application of ECs as BSM probes, we focus on 
a hadrophilic $Z^\prime$ gauge boson of mass $m_{Z^\prime}$ coupled to the SM quarks via
\begin{align}\label{Eq:Zprime}
    \mathcal{L}_{\text{int}} = \sum_q g'_{q} \bar{q}\slashed{Z'}q\,,
\end{align}
where $q$ runs over all SM quark flavors.\footnote{Strictly speaking, 
\eqref{Eq:Zprime} should be understood as an effective theory supplemented by 
a Wess--Zumino term, since the baryon number current is anomalous. 
See \cite{Dror:2017ehi,DiLuzio:2022ziu} for details.}
This model is the simplest realization of the two-prong resonance 
scenario discussed before, and its transverse production fixes the spinning coefficient to $c^{(2)}=1/2$ with all higher coefficients vanishing.

Beyond its simplicity, the hadrophilic $Z^\prime$ is a well-motivated and widely 
studied extension of the SM~\cite{Carone:1995pu,FileviezPerez:2010gw,Dobrescu:2013cmh,Ilten:2018crw}, 
making it a natural benchmark for our analysis. Practically, current bounds of order $g_q' \sim \mathcal{O}(0.1)$ allows large enough cross sections to look for resonant shapes of the form in Fig.~\ref{fig:resonantshape}.

We consider two production channels for a sufficiently light $Z'$ which gets boosted to be reconstructed as a single jet.
The first is an associated production with a photon, $pp \to Z' \gamma$, which 
provides a clean environment for a direct comparison with the state-of-the-art jet 
substructure techniques~\cite{CMS:2019xai}, and serves as our main 
setting to illustrate the key aspects of the analysis using ECs. 
Secondly, we discuss an associated production with a hard jet, 
$pp \to Z' j$ in the context of the existing CMS measurement of the 
ECs in dijet events~\cite{CMS:2024mlf} which provides with a broad view to realistic experimental and theory systematics.

\textit{\textbf{Associated production with a photon.---}}
We consider the production of the $Z'$ in association with a photon, following 
the CMS analysis~\cite{CMS:2019xai} from which we borrow the event 
selection and setup. The signal $pp \to Z'\gamma$ is generated at leading order (LO) in QCD with \MG, and
subsequently showered with \PT.  The nonresonant backgrounds, $\gamma$+jets, QCD multijet, and resonant ones $\gamma V$+jets ($V=W,\, Z$) are similarly simulated at LO by \MG\ with the default factorization and renormalization scales,
interfaced with the \PT. The NNPDF30 is used. All background samples were matched using $k_T$-jet MLM matching allowing up to four extra jets (two extra jets for $\gamma V$+jets background) in the 5-flavor scheme.

Events are required to contain exactly one photon with $p_{T\gamma} > 200\GeV$ and $|\eta_\gamma| < 2.4\,(2.1)$. 
All particles in an event are clustered into jets by {\tt{Fastjet}}~\cite{Cacciari:2005hq} using the anti-$k_T$ algorithm with a jet radius $R_j=0.8$ (AK8), and the leading jet is required to satisfy $p_{Tj} > 200\GeV$ and 
$|\eta_j| < 2.4$. This is taken as a baseline AK8 jet over which the inclusive EEC is computed. Further details on the CMS analysis exploiting the soft-drop mass $m_{\text{SD}}$, the shape variable $N_2$, and the related variables are reported in the Appendix. 

\begin{figure}
	\centering
	\includegraphics[width=0.98\columnwidth]{"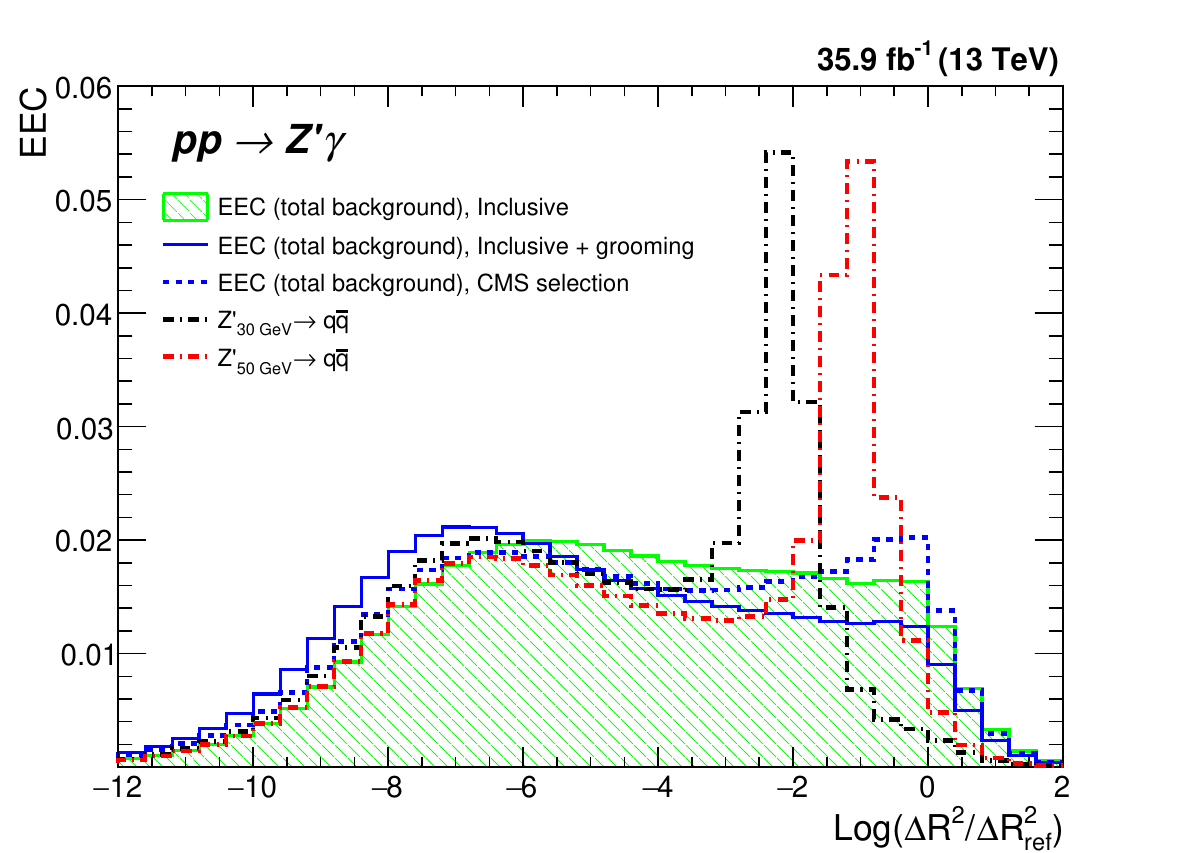"}
	\caption{The weighted average of EEC for two benchmark $Z'$ masses and total background for three different scenarios. Events are restricted to those with $p_{Tj} > 200$ GeV.}
	\label{fig:eecbkgCMS}
\end{figure}

Our data set and analysis are well validated by reproducing the soft drop mass distribution after imposing the same CMS selection, as detailed in the Appendix. The EEC is computed weighting by $p_{Ti}p_{Tj}/p_T^2$ over all pairs of particles inside a jet and then averaging over all the events. A comparison of the EEC spectra between the QCD background and signal is shown in Fig.~\ref{fig:eecbkgCMS}. The EEC is presented in terms of $\Delta R^2/\Delta R_{\text{ref}}^2$ where the reference scale is fixed in terms of the SM $Z$-boson mass, $\Delta R^2_{\text{ref}} = 4 m^2_Z/p^2_{Tj}$. 
In Fig.~\ref{fig:eecbkgCMS}, ECs originating from boosted $Z^\prime$ exhibit resonant shapes at $\Delta R^2_{\star} \sim 4 m^2_{Z^\prime}/p^2_{Tj}$, below which both background and signal spectra are similar, as one probes the individual quark jets from the $Z^\prime$ decay. At larger angles $\Delta R^2> \Delta R_{\star}^2$, the $Z^\prime$ spectrum rapidly falls off, as large-angle radiation is suppressed for color-singlet states.

While the cut on the shape variable $N_2$ in~\cite{CMS:2019xai} was crucial to reduce the background by the factor of 10, it introduces the artificial violation of the scaling as the cut retains only the phase space favorable for a two-prong resonance. The violation is clearly observed by comparing two curves (solid and dashed blue) in Fig.~\ref{fig:eecbkgCMS}, which differ mainly by the $N_2$ variable.\footnote{Although the grooming via the soft drop changes the overall distribution in Fig.~\ref{fig:eecbkgCMS}, compared to that from the baseline AK8 jet (shaded green), one sees that the scaling regions in both cases are well identified.}

For two-prong resonances of mass $m_X$, 
the ratio of angular distances scales as 
$\Delta R^2_{\star}/\Delta R_{\text{ref}}^2 \sim m^2_X/m^2_Z$, while its value is bounded by the jet radius as $\log(R_j^2/\Delta R_{\text{ref}}^2)$, which is roughly $-0.24$ for $p_{Tj} \sim 200$ GeV.  
Targeting a mass up to $m_{Z'} \sim 100$ GeV requires $p_{Tj}$ to be harder than $\sim 250$ GeV. 
To take into account the $p_T$ dependent behavior, events are split into $p_T$ bins in steps of 100 GeV up to 600 GeV and the overflow bin. 

To estimate the sensitivity to the $Z'$, we set up the $\chi^2$ test similarly to \cite{CMS:2024mlf}. For each $p_T$ bin, we compute the expected EEC in bins of $\log(\Delta R^2/\Delta R_{\text{ref}}^2)$ as a function of $g_q^\prime$ and $m_{Z^\prime}$. It is interesting to highlight that the expected product of fluxes at different bins is highly correlated, given in terms of a particular kinematical slice of the four-point correlator $\langle (\cale \cale)_i (\cale \cale)_j \rangle$, with $(\cale\cale)_k$ denoting the two-energy detectors separated by a distance in the $k$-th bin. Since we are using the same setup as \cite{CMS:2019xai} with sensitivity currently limited by statistics, we don't include any systematic uncertainty. We discuss this point below. The expected $95\%\text{CL}$ exclusion is computed from this $\chi^2$ following \cite{Cowan:2010js}, assuming that the measurement coincides with the SM expectation. 

\begin{figure}
	\centering
    \includegraphics[width=0.98\columnwidth]{"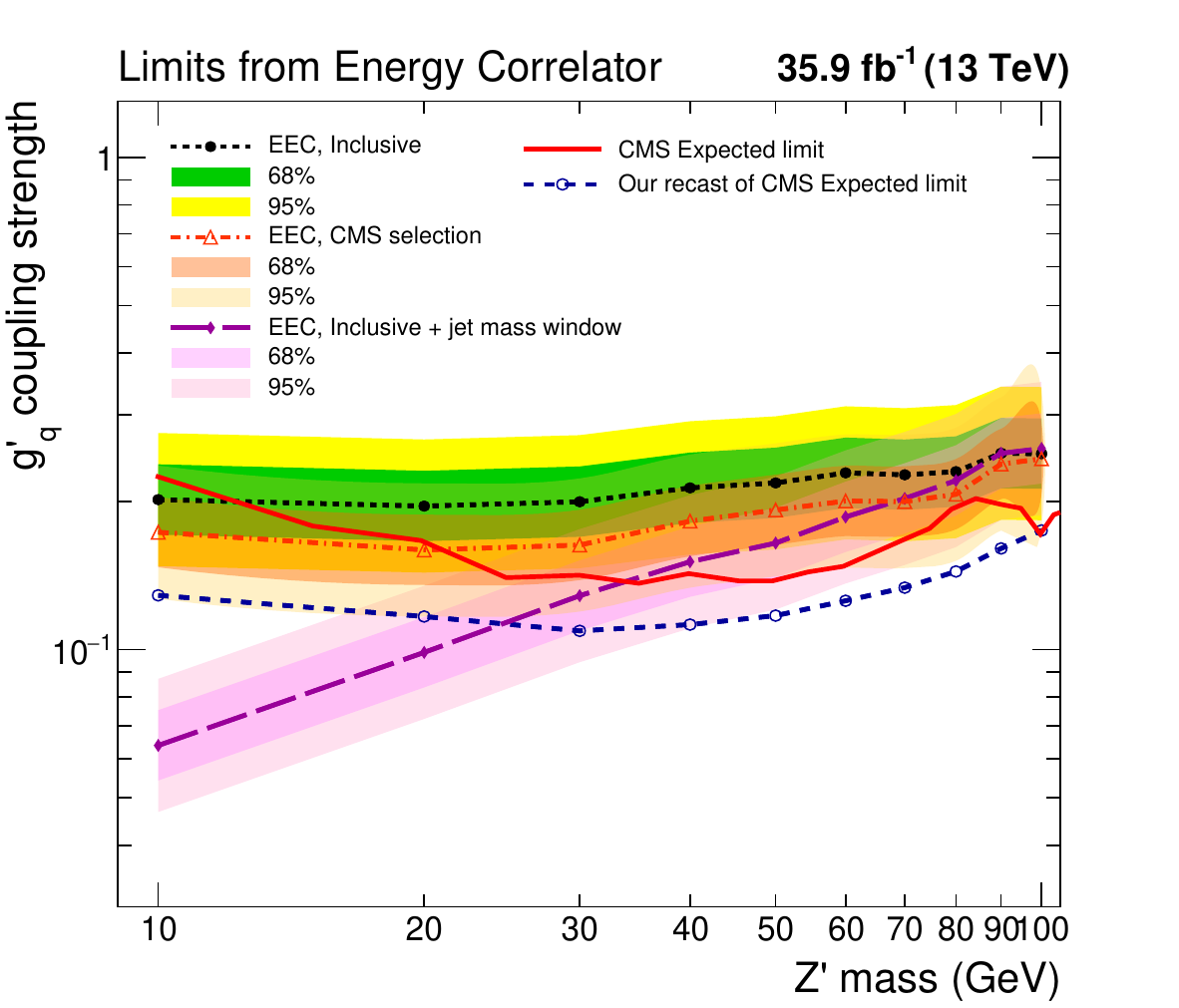"}
	\caption{Upper limits at 95\% CL on the coupling strength $g'_q$ of $Z'\rightarrow q\bar{q}$ from various EEC analyses and recasting of the CMS search on low mass resonances produced in association with a photon, $pp \rightarrow Z'\gamma$. The two bands for each limit represent 1 and 2 standard deviation intervals. 
     }
	\label{fig:photonjet_proj}
\end{figure}

The resulting sensitivity is presented in Fig.~\ref{fig:photonjet_proj}. The expected limit from the CMS Collaboration (solid red) was taken from~\cite{CMS:2019xai}. Our recast of the CMS bounds via the $m_{\text{SD}}$ measurement using our data is shown in dashed blue line. While it appears slightly stronger than the CMS result, we use our recast as the baseline to measure the quality of performances with EEC observables. The inclusive EEC (dotted black) in Fig.~\ref{fig:photonjet_proj} represents the sensitivity derived from the ungroomed baseline AK8 jets with no further cuts on their constituents. 
Compared to the inclusive EEC, the sensitivity from the EEC with AK8 jets on which the same CMS selection as in~\cite{CMS:2019xai} was imposed (dotted-dashed orange) does not show a meaningful improvement, as was indicated by the aforementioned violation of scaling. This implies that a precise measurement of the EEC at the inclusive level is roughly equivalent to imposing the CMS selection.  
Remarkably, the upper limit from the considerably simpler inclusive EEC observable is comparable with our CMS recast despite its factor of 10 bigger background size.

While being inclusive makes the inclusive EEC well suited for anomaly detection, once a more specific target is set combining it with some kinematic observables can be a way of enhancing the sensitivity or further improving in a particular parameter space. Given that the signal in terms of the jet mass is localized around $m_{Z^\prime}$, we adopt the (ungroomed) jet mass window ($\pm 20$\% of a $Z'$ mass) as the single selection in our inclusive EEC. 
Measuring the EEC in this jet mass window reveals a distinct behavior of the QCD background depending on the ratio $m_{j}/p_{Tj}$. While in the moderately boosted case the EEC gets considerably distorted around $\Delta R_{\star}^2$, in the highly boosted case, $m_{j}/p_{Tj} \ll 1$, the spectrum is similar to the inclusive case for $\Delta R^2<\Delta R_{\star}^2$. At larger angles the spectrum sharply decreases. In the same region, the signal still displays a sharp resonant feature, and therefore the measurement dramatically enhances the reach. More details are provided in the Appendix. 
We can see in Fig.~\ref{fig:photonjet_proj} how measuring inclusive EEC inside the jet mass window enhances the sensitivity (long-dashed pink) by reducing the background while keeping its smoothly falling shape. 

While we focused on the CMS analysis \cite{CMS:2019xai}, a similar discussion is expected to be applicable to the CMS search for the low-mass vector resonances~\cite{CMS:2017dcz} produced with ISR and the more recent ATLAS \cite{ATLAS:2024bms} analysis of the $j+\gamma$ channel.

\textit{\textbf{Learning from the CMS $\alpha_s$ measurement.--}}

The recent CMS measurement of ECs inside jets~\cite{CMS:2024mlf} precisely characterizes the energy-flow distributions in dijet events to extract the strong coupling $\alpha_S(m_Z)$. While not directly optimized for new physics searches, it provides direct access to realistic systematics in a well-defined experimental setup. As a proof of principle, we recast this analysis to derive limits on the light hadrophilic $Z'$ recoiling against a hard jet.

\begin{figure}
    \centering
    \includegraphics[width=1.05\linewidth]{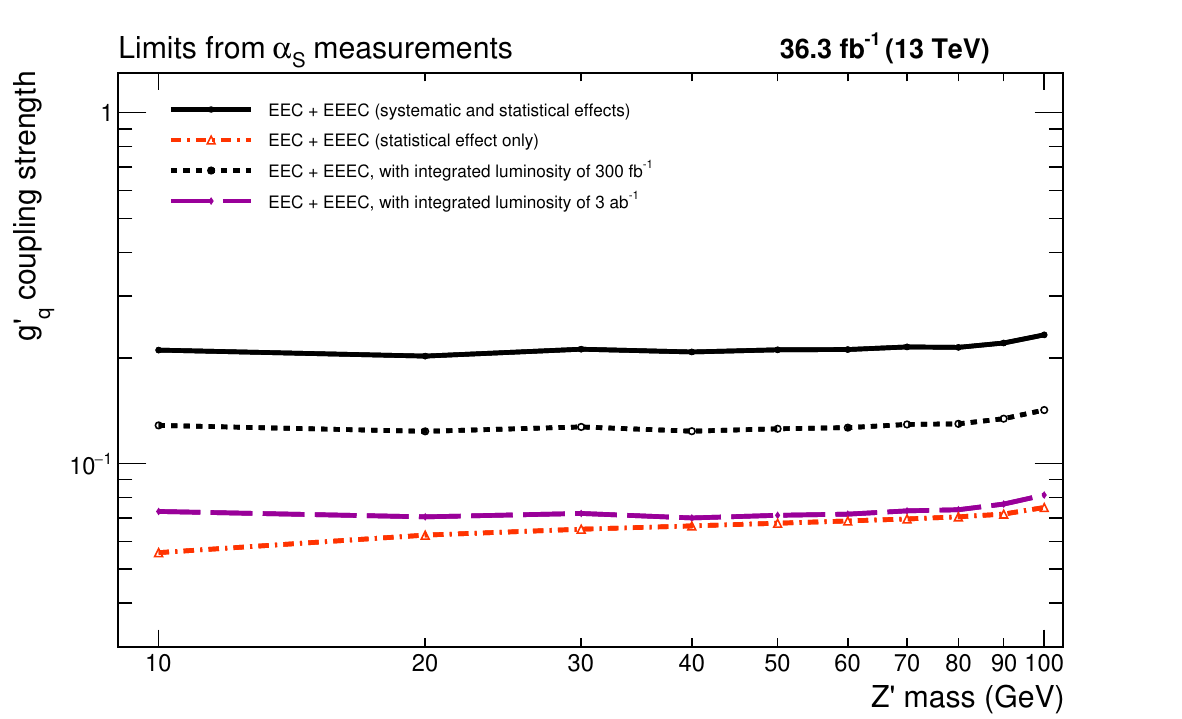}
    \caption{Projected limits at 95\% CL on the coupling-mass plane ($g_q',m_{Z'}$) of a hadrophilic $Z'$ derived from the extrapolation of the CMS energy correlator measurement in dijet events~\cite{CMS:2024mlf}. }
    \label{fig:Alphast}
\end{figure}

Following~\cite{CMS:2024mlf}, we simulate $Z'+\text{jet}$ signal events and compute the EEC and the three-point energy correlator (EEEC) on the leading AK4 jets in several $p_T$ windows; the EEEC is projected onto the longest angular side, integrated over the remaining angles. Full details on the event selection and observable construction are given in Appendix. The projected sensitivity is derived incorporating all sources of systematic and statistical uncertainties provided by CMS~\cite{CMS:2024mlf} and combining EEC and EEEC. The results are shown in Fig.~\ref{fig:Alphast}.

Interestingly, despite the event selection being far from optimal for a new
physics search, in particular the small clustering radius $R_j=0.4$, the
sensitivity is nonetheless competitive with other state-of-the-art searches,
as described above. Several remarks concern the systematics. The CMS analysis lists
9 sources of experimental systematic uncertainty, including scale variations, pileup, and
unfolding, and 6 sources of theoretical uncertainty, including PDF and $\alpha_s$ variations. In the CMS analysis \cite{CMS:2024mlf}, each systematic is assumed to be fully correlated among the bins, so that the covariance matrix of each systematic is $\Sigma^{ab} = \sigma^a \sigma^b$, with $a,b$ denoting the angular and $p_T$ bins (see the Appendix for more details). At the current luminosity of $ 36.3\,\text{fb}^{-1}$, these systematics are comparable in size to the statistical uncertainties and their inclusion has a noticeable impact on the reach, as shown in Fig.~\ref{fig:Alphast}.
However, the strong correlation implies that even if the
systematics do not improve, at higher luminosity it becomes increasingly
difficult for a new physics signal to be absorbed into systematic shifts, since it is the full shape of the EEC and EEEC distributions that drives the sensitivity. As
a result, the expected sensitivity improves with the luminosity, even under the assumption that the
systematics remain unchanged. We must emphasize, however, that our signal ECs are at particle level; detector granularity and smearing effects at these small angular scales, together with the unfolding procedure, may degrade the sensitivity compared to Fig.~\ref{fig:Alphast}. Nonetheless, these results encourage to reconsider these aspects in a dedicated experimental setup, with unfolded particle level treatment of systematics.  A concrete benchmark could be the measurement of the SM $Z,W$ boson contributions, which Fig.~\ref{fig:Alphast} suggests to be within reach.

\section{Outlook}\label{Sec:Outlook}

We proposed ECs as direct probes for new light resonances decaying to QCD. This work motivates several future directions. First, our results encourage a dedicated experimental and theoretical effort around EC-based BSM searches. As the $Z^\prime$ study demonstrates, EEC observables can reach sensitivity competitive with other dedicated searches, while exhibiting different systematics and maintaining a largely model-agnostic scope, making them a complementary approach to existing techniques. Furthermore, other classes of correlator-based observables, such as fully differential three-point correlator, energy$^n$ detectors or, when applicable, ratios of ECs between signal and control jets, may enhance the sensitivity. One may also explore a larger space of signals such as those producing multi-prong topologies and hidden strongly coupled scenarios, leading to nearly homogeneous energy distribution \cite{Ricci:2026qpw}.

As the LHC moves towards its high-luminosity phase, finding new observables becomes an important, prominent goal of phenomenology. As we investigated in this paper, ECs may provide a new way to repurpose precision jet substructure into novel BSM searches.

\subsection*{Acknowledgments}
We thank Bryan Cardwell, Heyjin Kwon, Kyle Lee, Ian Moult, Raman Sundrum, Jae Hyeok Yoo, Felix Yu, and Hua Xing Zhu for useful discussions. We are particularly grateful to Zhen Liu for collaboration during the initial stages of this project. We also thank the organizers and participants of Les Houches PhysTeV 2023, where this project was first discussed. L.R.\ is supported by the NSF grant PHY-2514660 and by the Maryland Center for Fundamental Physics. M.S.\ was supported by the National Research Foundation of Korea (NRF) under Grant Number RS-2024-00450835.

\bibliography{bibs}

\clearpage
\onecolumngrid

\begin{center}
\textbf{SUPPLEMENTAL MATERIAL}
\end{center}

\appendix
\setcounter{secnumdepth}{3}
\renewcommand{\thesection}{\arabic{section}}
\renewcommand{\thesubsection}{\thesection.\arabic{subsection}}
\renewcommand{\theequation}{\thesection.\arabic{equation}}
\makeatletter
\def\@appendixcntformat#1{\csname the#1\endcsname\quad}
\makeatother

\section{Validation against CMS analysis for low mass resonances with an associated photon}\label{App:cms:photon}

We add more details regarding our recast of the search for low-mass quark-antiquark resonances produced in association with a photon at $\sqrt{s}=13$ TeV~\cite{CMS:2019xai}.

Events are demanded to have only one photon (assuming 90\% identification efficiency) with $p_T(\gamma) > 200$ GeV
and $|\eta(\gamma)| < 2.4$ (2.1 for the final selection).
Events with additional photons with $p_T(\gamma) > 14$ GeV or additional leptons with $p_T(\ell) > 10$ GeV and $|\eta(\ell)|<2.5$ are vetoed.
Only events with the leading AK8 jet of $p_{Tj} > 200$ GeV and $|\eta(j)| < 2.4$ are considered. The leading AK8 jet and photon are required to have a minimal angular separation, $\Delta R(\gamma,\, p_{Tj}) > 2.2$.
The cut on the missing transverse momentum $p^\text{miss}_T <75$ GeV, which is efficient reducing the $t\bar{t}$ background (although we do not simulate this), is applied to all simulated backgrounds. Additionally, events with extra AK4 jets of $p_{Tj} > 30$ GeV found within $\Delta R = 0.8$ around the photon are vetoed. 

The QCD dijets matched up to two extra jets (we denote it as QCD multijet).  The QCD multijet events can radiate photons causing a double-counting. Any QCD events with an isolated photon (which should be part of $\gamma+$jets) are vetoed. To process QCD multijet samples, we first cluster every particle in an event into AK4 jets and take the leading AK4 jet of $p_T > 200$ GeV as the candidate faking a photon, and retain only events with no extra AK4 jets of $p_{Tj} > 30$ GeV within $\Delta R = 0.8$ around the fake photon. 
The fake rate of $j\rightarrow \gamma$ is applied a posteriori by applying the constant mistag rate $\mathcal{P}_{j\rightarrow \gamma}(p_T) =0.1\%$. 
The jet faking a photon is removed from the list and the remaining particles are re-clustered into AK8 jets. 
The leading AK8 jet of $p_{Tj} > 200$ GeV and $|\eta(j)|<2.4$ is taken as a $Z'$ candidate. 
All jets were smeared following the CMS performance study~\cite{CMS:2016lmd}, $\sigma_E/E = \sqrt{A^2/E^2 + B^2/E + C^2}$ where $A=5.0$, $B=1.0$, and $C=0.05$ were used. A discrepancy between the CMS estimate and ours caused by our simplified procedure on QCD multijet samples is subdominant as it is not a major contribution to the background. The $b$-jet vetoing is not applied as the $t\bar{t}$ background is not simulated.

The soft drop mass algorithm with $\beta = 0$ and $z_\text{cut} = 0.1$ (corresponding to the grooming) is used to remove the soft and wide-angle radiation from the jet, and the resulting groomed jet mass is labeled as $m_\text{SD}$.
The variable $\rho = \log (m^2/p_T^2)$ is useful in characterizing jets. It is roughly uncorrelated with the jet $p_T$ due to their own relation between jet mass and the jet $p_T$. Events are restricted to lie in the range, $ -7.5 < \rho  < -2.0$, 
where $\rho$ is evaluated with $m_\text{SD}$ and soft dropped jet $p_T$. Imposing cuts up to the above procedure defines the pre-selection.
\begin{figure}
	\centering
	\includegraphics[width=.58\columnwidth]{"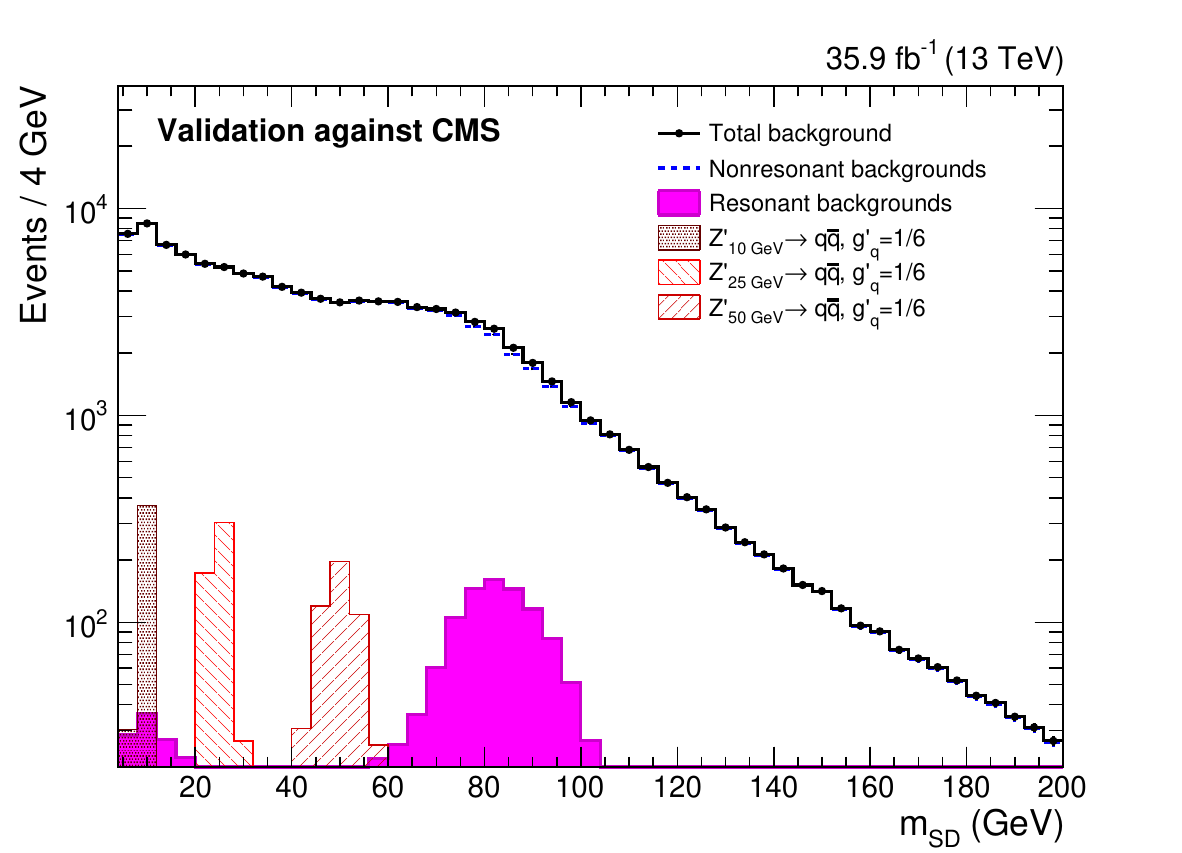"}
	\caption{Our validation against the CMS analysis after imposing the same set of cuts including the shape variable $N_2$  which retains only 10\% of the background.}
	\label{fig:validationCMS}
\end{figure}

The shape variable $N_2^1$ which is defined as (with different $\beta$ from that in the soft drop)
\begin{equation}
  N^\beta_2 = \frac{{}_2e^\beta_3}{({}_1e_2^\beta)^2}~,
\end{equation}
where
\begin{equation}
\begin{split}
  {}_1e_2^\beta = \sum_{i<j \,\in\, J} z_i\, z_j\, \Delta R_{ij}^\beta~, 
  \quad
  {}_2e_3^\beta= \sum_{i<j<k \,\in\, J} z_i\, z_j\, z_k
  \min\!\left\{\Delta R_{ij}^\beta \Delta R_{ik}^\beta,\;
               \Delta R_{ij}^\beta \Delta R_{jk}^\beta,\;
               \Delta R_{ik}^\beta \Delta R_{jk}^\beta\right\}~.
\end{split}
\end{equation}
$N_2^1$ is likely small for a two-prong decay. 
It is known that the fixed cut on $N^\beta_2 $ variable in $\gamma+$jets and QCD multijet samples has dependencies on the jet $\rho$ and  $p_T$ ($\rho$ is used as a proxy of the jet mass since it is roughly uncorrelated with the jet $p_T$), causing the distortion of the $m_\text{SD}$ distribution. To preserve the shape of the $m_\text{SD}$ distribution, a varying cut on $N^1_2$ as a function of $(\rho^\text{jet},\, p^\text{jet}_T )$ is introduced, namely $N^\text{DDT}_2 (\rho^\text{jet},\, p^\text{jet}_T ) = N^1_2 - X_{10\%} (\rho^\text{jet},\, p^\text{jet}_T )$ where $X_{10\%}$ is the value that retains only 10\% of the background at each $(\rho^\text{jet},\, p^\text{jet}_T )$. 
Events are demanded to satisfy $N^\text{DDT}_2 (\rho^\text{jet},\, p^\text{jet}_T ) <0$.
For simplicity, in our validation analysis, we obtain one-dimensional projected $N_2$ distribution of the non-resonant background, namely $\gamma$+jets and QCD multijet, for each $(\rho^\text{jet},\, p^\text{jet}_T )$ bin and determine $X_{10\%}$ that keeps only 10\% of total events in the corresponding bin. We repeat the same exercise over all $N_{\rho} \times N_{p_T}$ bins in ranges $p_{Tj} = [200,\, 1000]$ GeV and $\rho = [-7.5,\, -2]$,  leading to the matrix elements $(X_{10\%})_{ij}$, and it is applied to estimate the final number of total signal plus background events. 
$N_{\rho} = N_{p_T}$ = 10 was used.
The final validation for the soft drop jet mass distribution after imposing all cuts including $N^2_\text{DDT} < 0$ is shown in Fig.~\ref{fig:validationCMS}. It shows quite a good agreement with the CMS result in~\cite{CMS:2019xai} which strongly validates our procedure.

\section{More on combining inclusive EEC with jet mass window}

Previously, we demonstrated the impact on the sensitivity from the inclusive EEC combined with additional kinematic variable, such as the jet mass window, in the process $pp \rightarrow Z'\gamma$ at the LHC. 

\begin{figure}[h]
	\centering
    \includegraphics[width=.48\columnwidth]{"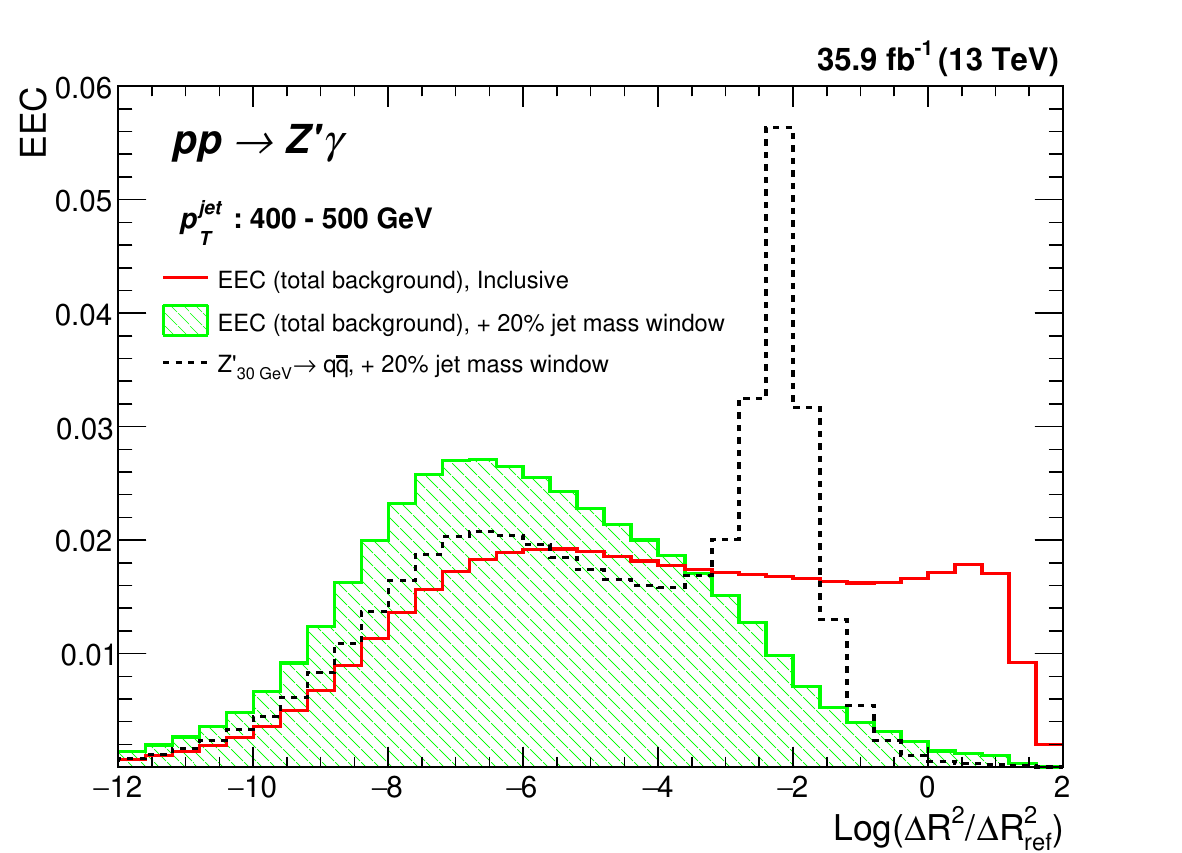"}
	\caption{EEC distributions of the signal and backgrounds in $p_T = [400,\, 500]$ GeV with and without the jet mass window $m_j = 30$ GeV $\pm$ 20\%. }
	\label{fig:EEC:jetmasswindow:sig}
\end{figure}

\begin{figure}[h]
	\centering
    \includegraphics[width=.48\columnwidth]{"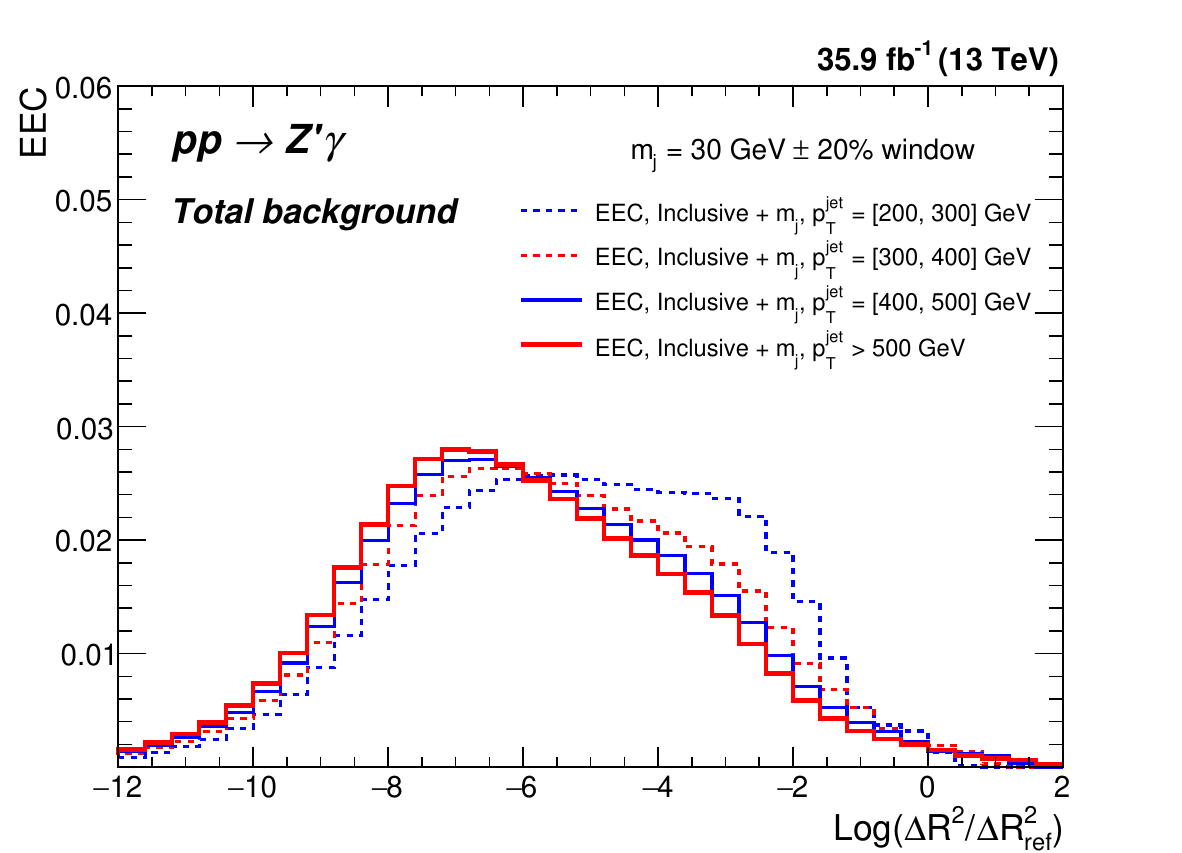"}
    \includegraphics[width=.48\columnwidth]{"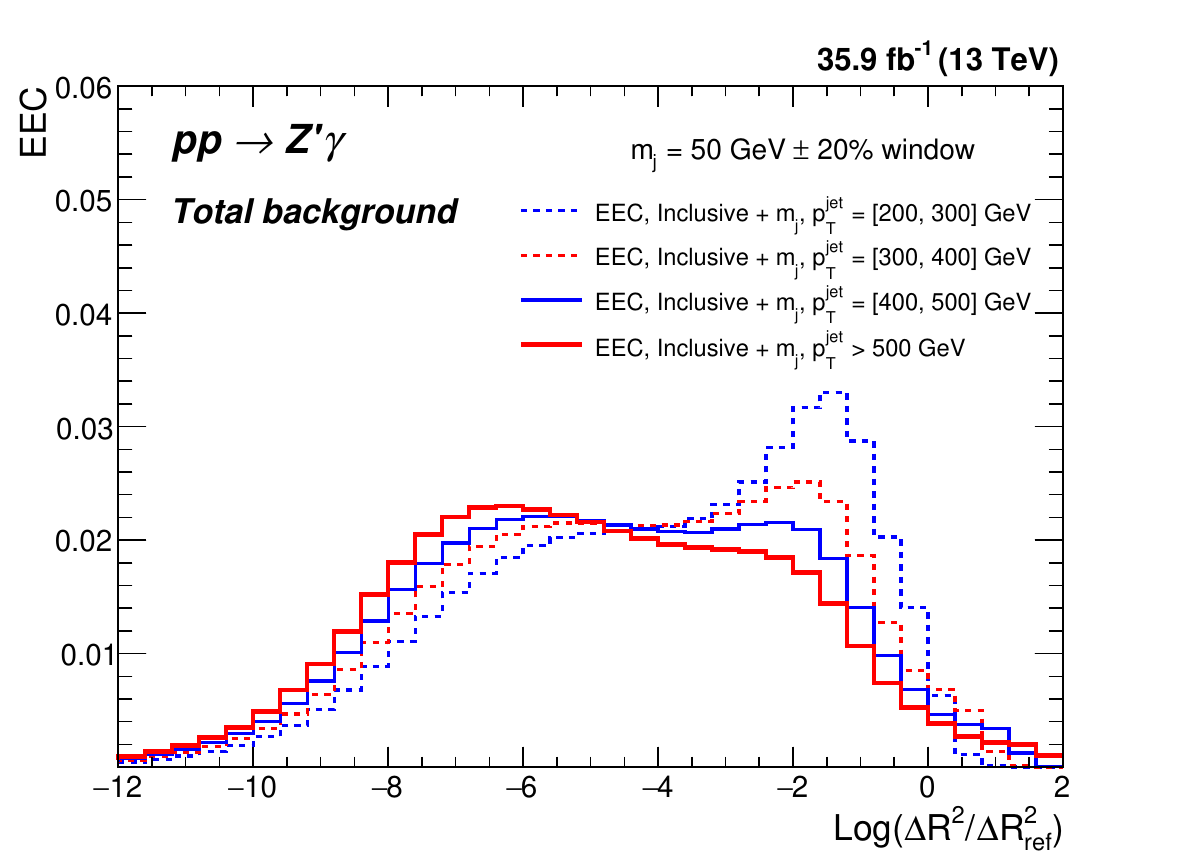"}
	\caption{EEC distributions of total backgrounds in various $p_T$ bins for two benchmark jet mass windows $m_j = 30$ GeV $\pm$ 20\% (left) and $m_j = 50$ GeV $\pm$ 20\% (right). }
	\label{fig:EEC:jetmasswindow:bkg}
\end{figure}

In Fig.~\ref{fig:EEC:jetmasswindow:sig}, we display the EEC of the signal on top of two EEC distributions of the total backgrounds differing by only jet mass window, $m_{j} = 30$ GeV $\pm$ 20\%, in the $p_T$ window $p_T = [400,\, 500]$ GeV. In Fig.~\ref{fig:EEC:jetmasswindow:bkg}, we illustrate how the EEC distribution of the total background changes as the phase space transit from a moderately boosted region to the highly boosted one, $m_j/p_{Tj} \ll 1$ for two benchmark mass windows, $m_j = 30$ GeV $\pm$ 20\% and 50 GeV $\pm$ 20\%. 
As is evident in Fig.~\ref{fig:EEC:jetmasswindow:bkg}, restricting to the phase space within the jet mass window for relatively low values of $p_T$ introduces a pronounced artificial shape due to the correlation of the jet mass to the angular distance. However, increasing $p_T$, while the  jet mass window is fixed, suppresses the artificial shape around the peak position of the signal and thus retains the smoothly falling behaviour with lower overall background rate. Since jets are required to be increasingly boosted as the jet mass grows, the improvement is more pronounced in the low-mass region.

\section{Learning from the CMS $\alpha_s$ measurement}\label{App:alpha}

Here we provide details on the sensitivity projections of Fig.~\ref{fig:Alphast} based on the CMS measurement of two- and three-point energy correlators in dijets~\cite{CMS:2024mlf}.

\textit{\textbf{Event selection.}}---
We select dijet events at $\sqrt{s}=13~\mathrm{TeV}$.
Jets are clustered with the anti-$k_T$ algorithm ($R=0.4$) from final-state particles, excluding neutrinos and charged leptons but including photons, with a constituent $p_T$ threshold of $1~\mathrm{GeV}$.
An event is accepted if it contains at least two jets satisfying
\begin{equation}
  p_T > 97~\mathrm{GeV}, \qquad |\eta| < 2.1, \qquad |\Delta\phi(j_1,j_2)| > 2.0~\mathrm{rad}.
\end{equation}
The energy correlators are computed on the leading two jets.
Jets are placed into seven $p_T$ bins (in $\mathrm{GeV}$):
$[220,330]$, $[330,468]$, $[468,638]$, $[638,846]$, $[846,1101]$, $[1101,1410]$, $[1410,1784]$; we drop the lowest $p_T$ bin provided by CMS.
A single event may contribute to two different bins if the leading jets fall in different $p_T$ ranges.

\textit{\textbf{Observables.}}---
For a jet with total energy $E_\mathrm{jet}$ and constituents indexed by $i$,
define the energy fraction $z_i = E_i/E_\mathrm{jet}$.
The two- and three-point energy correlators are projected onto the longest angular side $x_L$:
\begin{align}
  \mathrm{EEC}(x_L) &= \sum_{i,j}\, z_i\, z_j\;
    \delta\!\left(x_L - \Delta R_{ij}\right), \\
  \mathrm{EEEC}(x_L) &= \sum_{i,j,k}\, z_i\, z_j\, z_k\;
    \delta\!\left(x_L - \max\{\Delta R_{ij},\Delta R_{ik},\Delta R_{jk}\}\right),
\end{align}
where $\Delta R_{ab}=\sqrt{(\Delta\eta_{ab})^2+(\Delta\phi_{ab})^2}$.
Both distributions are binned in $x_L$ using the 23 edges
\begin{align}
  \label{Eq:xLBins}
  \{&0,\ 0.0002,\ 0.001,\ 0.002,\ 0.005,\ 0.01,\ 0.02,\ 0.04,\ 0.06,\ 0.08,\ 0.1,\ 0.12,\nonumber\\
    &0.137,\ 0.157,\ 0.179,\ 0.205,\ 0.234,\ 0.268,\ 0.306,\ 0.35,\ 0.4,\ 0.8,\ 1.0\},
\end{align}
normalized by the number of jets per $p_T$ bin. We restrict the analysis to $0.002 \leq x_L \leq 0.35$, excluding the extremal bins to avoid boundary effects. Note that the CMS $\alpha_s$ extraction~\cite{CMS:2024mlf} uses a tighter range, $20\,\mathrm{GeV}/p_{T,j} \lesssim x_L \lesssim 0.2$, retaining only the fully perturbative region well within the jet boundary. In contrast, we adopt a wider range to maximize sensitivity, at the cost of a less controlled theoretical description; the associated theory systematics are included in the $\chi^2$ as described below.

\textit{\textbf{Signal injection.}}---
Signal and background events are generated at leading order with \MG\ and showered with \PT.
Since CMS provides only the shapes of the background distributions, we inject the signal as
\begin{align}
  \mathcal{P}_{\text{CMS}}(x_L)
  \;\longrightarrow\;
  \mathcal{P}_{\text{CMS}}(x_L)
  + \frac{N^{\text{ev}}_{\text{BSM}}}{N^{\text{ev}}_{\text{SM}}}
    \mathcal{P}_{\text{BSM}}(x_L),
\end{align}
where $\mathcal{P}_{\text{BSM}}$ is the normalized BSM distribution and $N^{\text{ev}}_{\text{SM,BSM}}$ are the numbers of simulated events. The background distributions enter only through the event-count ratio and never appear directly in the $\chi^2$. To avoid statistical correlations between the two observables, the signal sample is split equally between the EEC and EEEC analyses, following~\cite{CMS:2024mlf}.

\begin{figure}
	\centering
	\includegraphics[width=.48\columnwidth]{"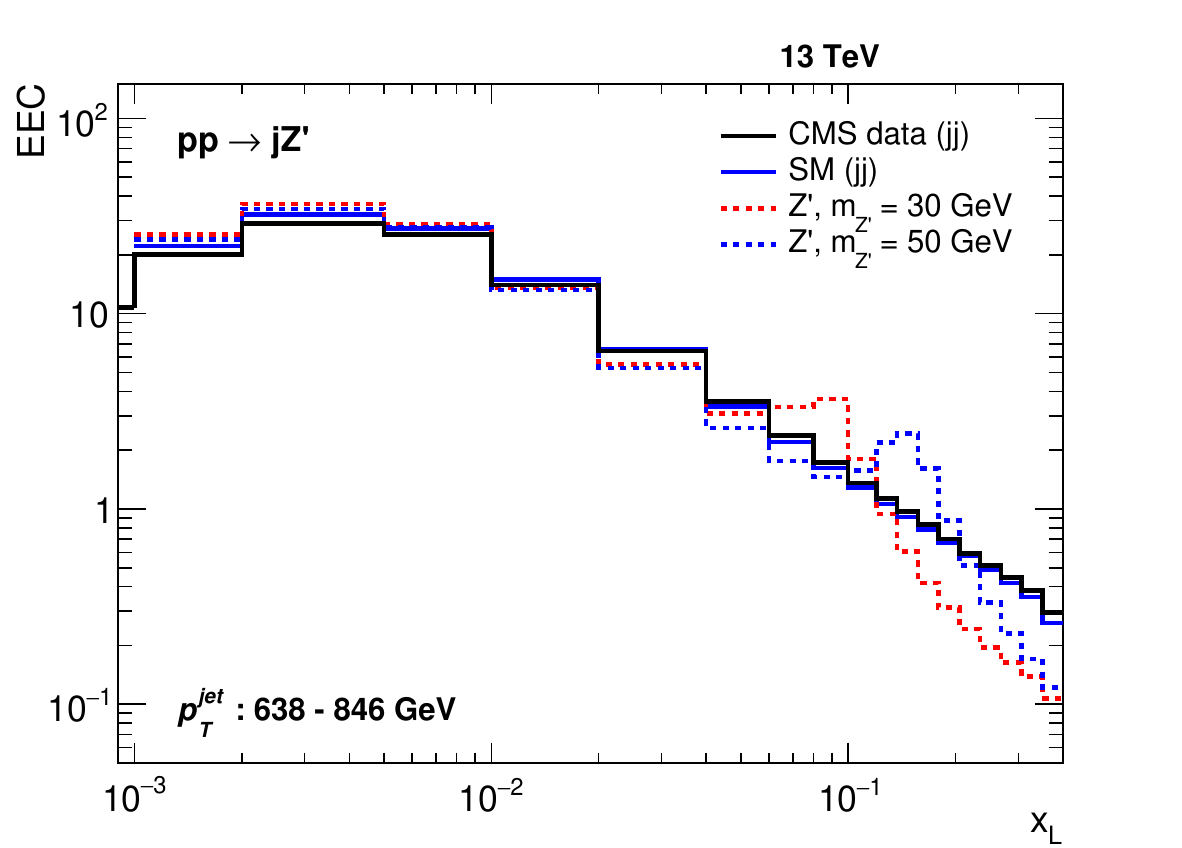"}
    \hfill
    \includegraphics[width=.48\columnwidth]{"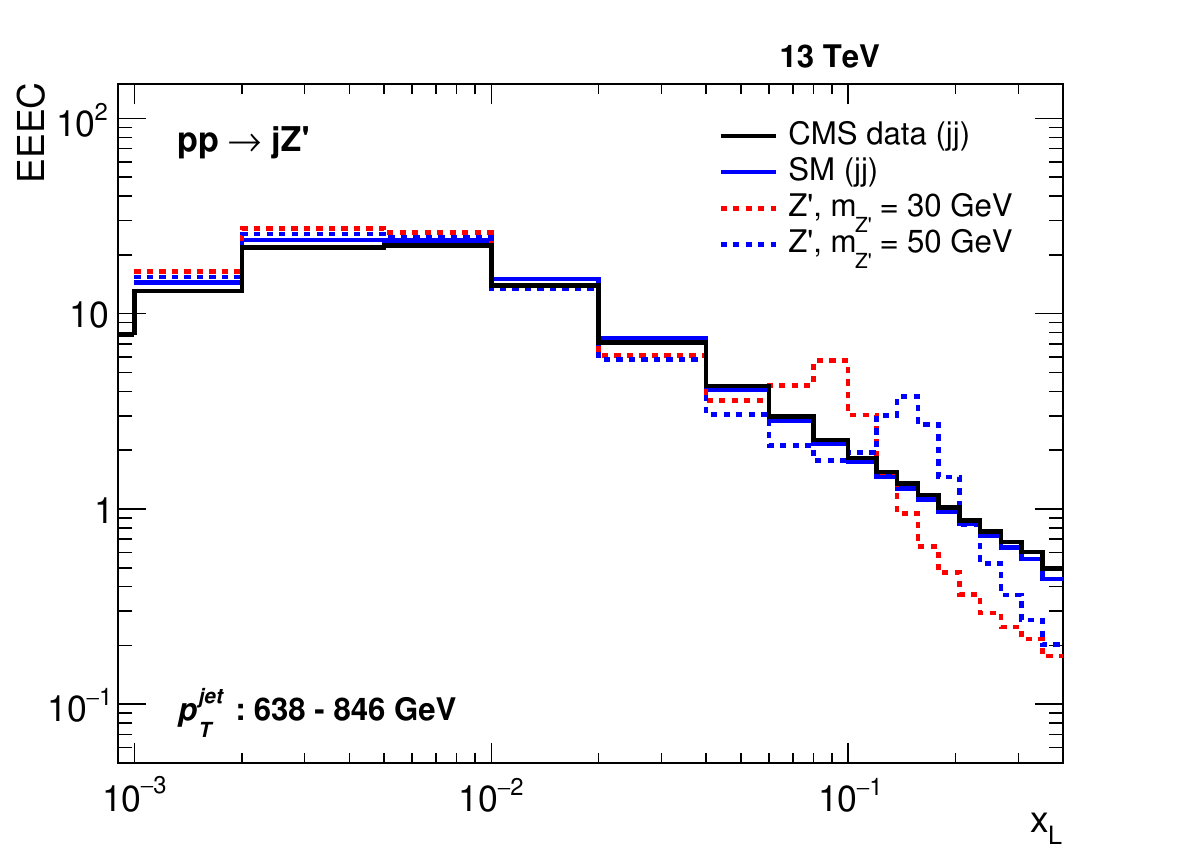"}
	\caption{EEC (left) and EEEC (right) distributions as a function of $x_L$ for QCD background and injected $Z'$ signal at representative benchmark masses, in a selected $p_T$ bin. The CMS data are taken from \cite{CMS:2024mlf}.}
	\label{Fig:EC_alpha}
\end{figure}

Representative EEC and EEEC signal shapes for selected benchmark masses and $p_T$ bins are shown in Fig.~\ref{Fig:EC_alpha}.

\textit{\textbf{Statistical analysis.}}---
The sensitivity is quantified via
\begin{align}\label{Eq:Chi2}
  \chi^2(g_q', m_{Z'})
  = \sum_{\substack{p_T,x_L \\ p_T',x_L'}}
    \Delta\mathcal{P}(p_T,x_L)\,
    \bigl[\Sigma_{\text{exp}} + \Sigma_{\text{TH}} + \Sigma_{\text{stat}}\bigr]^{-1}_{(p_T,x_L),(p_T',x_L')}\,
    \Delta\mathcal{P}(p_T',x_L'),
\end{align}
where $\Delta\mathcal{P} = (N^{\text{ev}}_{\text{BSM}}/N^{\text{ev}}_{\text{SM}})\,\mathcal{P}_{\text{BSM}}$ is the injected BSM excess, and we take the data to coincide with the central SM prediction (Asimov dataset). Each asymmetric uncertainty source enters with a symmetrized value $\sigma = (\sigma^+ + \sigma^-)/2$. The covariance matrices are
\begin{align}
  \Sigma_{\text{X}}(p_T,x_L;\,p_T',x_L')
  &= \sum_{i \in \text{X}}
     \sigma^{i}(p_T,x_L)\,
     \sigma^{i}(p_T',x_L'), \qquad \text{X} = \text{exp, TH}, \\
  \Sigma_{\text{stat}}(p_T,x_L;\,p_T',x_L')
  &= \sigma_{\text{stat}}(p_T,x_L)\,
     \Sigma_{\text{corr}}(p_T,x_L;\,p_T',x_L')\,
     \sigma_{\text{stat}}(p_T',x_L').
\end{align}

For the $300\,\text{fb}^{-1}$ and $3\,\text{ab}^{-1}$ projections, $\Sigma_{\text{stat}}$ is rescaled by the corresponding luminosity ratio relative to the baseline $36.3\,\text{fb}^{-1}$, while the systematic covariances are kept unchanged.

The nine experimental systematic sources are: jet energy scale (JES), jet energy resolution (JER), pileup, prefiring, charged-particle energy scale, neutral-particle energy scale, photon energy scale, track efficiency, and unfolding model (\MG+\PT).
The six theoretical systematic sources are: QCD scale of the parton shower, QCD scale of the hard scattering, tune, CNS, PDF, and PDF $\alpha_s$.
The $\chi^2$ values from the EEC and EEEC analyses are combined to derive the final sensitivity.

We note that in the CMS $\alpha_s$ analysis the fitting range is restricted to $20 \GeV/p_{T,j} \lesssim x_L < 0.2$, where the EEEC/EEC ratio is under full analytic control and theory systematics are negligible. Unfortunately, this upper cut excludes a large fraction of the signal, which motivated our choice of the wider range $x_L \lesssim 0.35$ at the cost of larger theory systematics, as discussed above. With a larger jet radius and an extended perturbative window, the EEEC/EEC ratio could in principle be exploited for new physics searches with reduced systematics, and it would be interesting to explore this possibility. Finally, we also verified the robustness of our results by excluding the two highest-$p_T$ bins, where the angular peak associated with most of the signal mass range is pushed to very small angles and may be less well resolved. We find only a marginal deterioration with respect to Fig.~\ref{fig:Alphast}.

\end{document}